\documentclass[english,aps,prl,twocolumn]{revtex4-1}
\usepackage[T1]{fontenc}
\usepackage[latin9]{inputenc}
\usepackage{float}
\usepackage{graphicx}
\usepackage{amsmath,amssymb}
\usepackage{esint}
\usepackage{color}

\makeatletter
\@ifundefined{textcolor}{}
{%
 \definecolor{BLACK}{gray}{0}
 \definecolor{WHITE}{gray}{1}
 \definecolor{RED}{rgb}{1,0,0}
 \definecolor{GREEN}{rgb}{0,1,0}
 \definecolor{BLUE}{rgb}{0,0,1}
 \definecolor{CYAN}{cmyk}{1,0,0,0}
 \definecolor{MAGENTA}{cmyk}{0,1,0,0}
 \definecolor{YELLOW}{cmyk}{0,0,1,0}
 }

\makeatother

\usepackage{babel}
\begin{document}
\global\long\def\bra#1{\left\langle #1\right|}

\global\long\def\ket#1{\left|#1\right\rangle }

\global\long\def\bk#1#2#3{\bra{#1}#2\ket{#3}}

\global\long\def\ora#1{\overrightarrow{#1}}

\title{Gate controlled Spin-Density Wave and Chiral FFLO Superconducting phases in interacting  Quantum Spin Hall edge states}

\author{Qinglei Meng$^{1}$, Taylor L. Hughes$^{1}$, Matthew J. Gilbert$^{2,3}$, Smitha Vishveshwara$^{1}$}

\affiliation{$^1$Department of Physics, University of Illinois, 1110 West Green St, Urbana IL 61801}
\affiliation{$^2$Department of Electrical and Computer Engineering, University of Illinois, 1406 West Green St, Urbana IL 61801}
\affiliation{$^3$Micro and Nanotechnology Laboratory, University of Illinois, 208 N. Wright St, Urbana IL 61801}
\date{\today}

\begin{abstract}
We explore the phases exhibited by  an interacting quantum spin Hall edge state in the presence of finite chemical potential (applied gate voltage) and spin imbalance (applied magnetic field). We find that the helical nature of the edge state gives rise to orders that are expected to be absent in non-chiral one-dimensional electronic systems. For repulsive interactions, the ordered state has an oscillatory spin texture whose ordering wavevector is controlled by the chemical potential. We analyze the manner in which a magnetic impurity provides signatures of such oscillations. We find that finite spin imbalance favors a finite current carrying groundstate that is not condensed in the absence of interactions and is superconducting for attractive interactions. This state is characterized by FFLO-type oscillations where the Cooper pairs obtain a finite center of mass momentum. 
\end{abstract}

\maketitle

The quantum spin Hall (QSH)/2D time-reversal invariant topological insulator state\cite{Kane2005a,Kane2005b,bernevig2006a,Taylor2006,koenig2007} has been attracting much interest since its recent discovery. The bulk of this insulating state is gapped, and characterized by a non-trivial $Z_2$ topological invariant which distinguishes this state from an ordinary band insulator. The most interesting low energy physics occurs at the QSH edge where an odd-number of counter-propagating Kramers' pairs of edge states exist. These edge states form a \emph{helical} liquid wherein the spin polarization is correlated with the direction of motion. Notably, these counter-propagating channels are robust even when disorder is present as long as time-reversal symmetry is preserved. This new type of 1D liquid can exhibit  unusual  features, such as, fractional charge, Kramers' pairs of Majorana bound states, and individual Majorana bound states when in the proximity of magnets, superconductivity, or both, respectively
\cite{qi2008, Fu2008, Vasudha2010}.

In this work, instead of considering proximity-induced effects in the QSH edge, we explore the rich 1D phases intrinsic to the helical liquid in the presence of interactions.
 We observe that repulsive interactions can lead to a spin-density wave phase that is unique to the helical liquid and argue that it is not generated in a typical 1D electron gas (1DEG). On the other hand, attractive interactions render the liquid unstable to the formation of a  Fulde-Ferrell-Larkin-Ovchinnikov (FFLO)-type superconducting phase\cite{FF1964,LO1964}, which is an s-wave like order parameter that condenses at a finite wave-vector.   The 1D nature of the helical liquid gives it an upper hand for hosting the FFLO-type phase, which has eluded experimental observation in higher dimensions\cite{Yang2001,zhao2008,Erich2010}. The two phases emerging from repulsive versus attractive interactions are dual to each other in that the roles of spin and charge are exchanged. Finally, we show that with repulsive interactions a magnetic impurity acts as an effective experimental probe of the QSH edge, inducing 
 oscillations in the magnetization \emph{direction} which are
fundamentally different from the Ruderman-Kittel-Kasuya-Yosida (RKKY) oscillations of the magnetization \emph{amplitude}
in a conventional 1DEG\cite{Ruderman1954,Kasuya1956,Yosida1957}.

\begin{figure}[t]
\centering{}\includegraphics[width=0.48\textwidth]{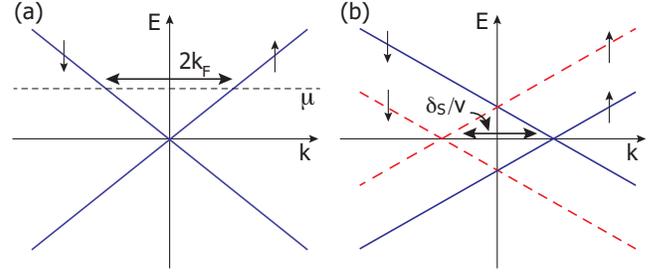}\caption{\label{fig:banddiagram}Non-interacting picture for spin density wave and chiral FFLO-type superconductor state formation (a)Edge state energy spectra at chemical potential $\mu.$ It is energetically favorable to open a gap at the Fermi-level as opposed to the Dirac point, thus forming a spin-density wave with wavevector $2k_F.$ (b) Bogoliubov-de Gennes spectrum for non-zero $\delta_S.$ Solid lines are electron states, dashed lines are hole states. Hybridization must occur between a solid and dashed line with opposite spin leading to a finite pairing wavevector of $\delta_S/v.$}
\end{figure}

We begin with a heuristic analysis of the non-interacting QSH edge in the presence of finite chemical potential and spin imbalance, focusing on the fundamental differences that give rise to new order when compared with a typical 1DEG. As shown in  Fig. \ref{fig:banddiagram}a, the QSH edge consists of linearly dispersing spin-dependent modes associated with a Dirac point centered at zero momentum, and is described by the Hamiltonian
\begin{eqnarray}
H_{0}&=&v\int dx \left[\psi_{R\uparrow}^{\dagger}(x)(-i\partial_{x})\psi_{R\uparrow}(x)\right.\nonumber\\&-&\left. \psi_{L\downarrow}^{\dagger}(x)(-i\partial_{x})\psi_{L\downarrow}(x)\right]\label{eq:QSH}
\end{eqnarray}
where $v\ $is the velocity and $x\ $is the coordinate tangent to the edge of the QSH sample. The operator $\psi_{R\uparrow(L\downarrow)}(x)$ annihilates electron moving to the right(left) with up(down) spin at position
$x$. The effects of a non-zero chemical potential and a spin imbalance can be described by
\begin{equation}
H_{\mu}=\int dx(-\mu_{\uparrow}\psi_{R\uparrow}^{\dagger}(x)\psi_{R\uparrow}(x)-\mu_{\downarrow}\psi_{L\downarrow}^{\dagger}(x)\psi_{L\downarrow}(x)),
\end{equation}
where $\mu_{\uparrow(\downarrow)\ }$is an effective chemical potential for up (down) spin in the helical liquid. The chemical potential,  $\mu=\frac{1}{2}(\mu_{\uparrow}+\mu_{\downarrow})$, can be controlled by tuning a gate voltage, and the spin imbalance,  $\delta_S=\mu_{\uparrow}-\mu_{\downarrow}$, may be controlled by applying magnetic field in the direction perpendicular to the QSH plane (or more generally, parallel to the direction of spin-polarization of the edge state). In fact, because of the spin-momentum locking on the edge, a \emph{spin} imbalance acts to give rise to a \emph{charge} current.

Given the fundamental fields comprising the QSH edge,  the two lowest order operators that could develop non-vanishing expectation values in an ordered phase are 
\begin{equation}
O_{m}=\psi_{R\uparrow}^{\dagger}(x)\psi_{L\downarrow}(x),\ \ \ O_{\Delta}=\psi_{R\uparrow}(x)\psi_{L\downarrow}(x),\label{eq:m}
\end{equation}
These order parameters represent magnetic order ($\langle O_m \rangle$) and superconducting order ($\langle O_\Delta \rangle$) and are dual to one another with regards to charge and spin in that the former carries charge $0$ and spin $2\hbar/2$ while  $\langle O_m\rangle$ carries charge $2e$ and spin $0$. 

We now argue that for non-vanishing $\mu$ and $\delta_S$, these order parameters have the striking property that they are inhomogeneous in space, exhibiting oscillatory behavior over a characteristic length scale.  We begin by tuning $\mu=\delta_{S}=0$ and considering magnetic order.  The system is tuned to the Dirac point and and any ferromagnetic order perpendicular to the spin-polarization of the edge states would open a gap at $k=0$ since it would couple via a constant multiplying a Pauli spin matrix. If one tunes $\mu$ away from zero then, in order to open a gap at the Fermi-level, the magnetic order must have a finite wave-vector of $q^{(0)}_m\equiv -2\mu/v=-2k_F$ where the superscript refers to the free limit(see Fig. \ref{fig:banddiagram}a). Thus, we induce a spin-density wave so that a gap can open at the Fermi-level as opposed to a gap opening at the (buried) Dirac point for ferromagnetic ordering. This type of chemical potential driven spin-density wave is unique to the helical liquid as seen by noting the form of a magnetic order parameter for a full 1DEG:
\begin{eqnarray}
&\psi^{\dagger}_{\uparrow}(x)\psi^{\phantom{\dagger}}_{\downarrow}(x)&\; \sim \left(e^{-ik_F x}\psi^{\dagger}_{R\uparrow}(x)+e^{ik_{F}x}\psi^{\dagger}_{L\uparrow}(x)\right)\nonumber\\ &&\times \left(e^{ik_F x}\psi^{\phantom{\dagger}}_{R\downarrow}(x)+e^{-ik_{F}x}\psi^{\phantom{\dagger}}_{L\downarrow}(x)\right)\nonumber\\
&=&\psi^{\dagger}_{R\uparrow}\psi^{\phantom{\dagger}}_{R\downarrow}+\psi^{\dagger}_{L\uparrow}\psi^{\phantom{\dagger}}_{L\downarrow}+\left(e^{-2ik_F x}\psi^{\dagger}_{R\uparrow}\psi^{\phantom{\dagger}}_{L\downarrow}+\textrm{c.c.}\right).\nonumber
\end{eqnarray}\noindent For the full 1DEG the non-oscilliatory terms generically dominate, but these terms are completely absent for the helical liquid which only has $e^{-2ik_Fx}\psi^{\dagger}_{R\uparrow}\psi^{\phantom{\dagger}}_{L\downarrow}$ non-vanishing. Thus, the existence of a spin-density wave is a unique signature of the reduced degrees of freedom of the helical liquid as compared to a conventional 1DEG.

Now let us consider the effects of a non-zero $\delta_S$ in the non-interacting limit for which we will return to the free-fermion Hamiltonian.  In the Bogoliubov-de Gennes formalism the Hamiltonian can be re-written
\begin{eqnarray}
H_{BdG}=\int dx \Psi^{\dagger}(x)\left(-iv\partial_x\mathbb{I}\otimes\sigma^z-(\delta_{S}/2)\tau^z\otimes\sigma^z\right)\Psi(x)\nonumber
\end{eqnarray}\noindent 
where $\tau^a$ represents particle-hole space and $\sigma^a$ spin space, and $\Psi(x)=(\psi_{R\uparrow}(x)\;\; \psi_{L\downarrow}(x)\;\; \psi^{\dagger}_{R\uparrow}(x)\;\;\psi^{\dagger}_{L\downarrow}(x))^T.$ This has energy levels $E_{\pm}=\pm\vert vk\vert \pm \delta_{S}/2$ (with uncorrelated signs). A homogenous s-wave pairing cannot open a gap at the Fermi-level if $\delta_S\neq 0$ and is thus energetically frustrated. As shown in Fig. \ref{fig:banddiagram}b the pairing term must have a finite wave-vector $q^{(0)}_{\Delta}\equiv \delta_{S}/v$ in the non-interacting limit to open a gap. A full 1DEG would have both $\psi^{\dagger}_{R\uparrow}\psi^{\dagger}_{L\downarrow}+\psi^{\dagger}_{R\downarrow}\psi^{\dagger}_{L\uparrow}$ pairing terms while the QSH edge only has the former. Thus, in the helical case there is always a ground state current of Cooper pairs in one direction picked by the sign of $\delta_S$ since the order parameter oscillates like $e^{iq_{\Delta}x}$ instead of $\cos (q_{\Delta}x)$, as in fact originally considered by Fulde and Ferrell\cite{FF1964} .We refer to this state as a chiral FFLO state.

We now turn to the effects of interactions and their crucial role in determining the fate of the QSH edge state and  energetically favorable ordered states. We derive the corresponding phase diagram by analyzing the form of the susceptibilities associated with each order and show, as might be expected, that magnetic (superconducting) order is stabilized by repulsive (attractive) interactions. We note that we are only considering equilibrium states. As in previous treatments\cite{Wu2006,Xu2006,Kim2009}, we ignore Umklapp scattering and employ the following form for interactions between QSH electrons:
\begin{eqnarray}
H_{I} & = & \frac{g_{4}}{2}\int dx\psi_{R\uparrow}^{\dagger}(x)\psi_{R\uparrow}(x)\psi_{R\uparrow}^{\dagger}(x)\psi_{R\uparrow}(x)\nonumber \\
 & + & \frac{g_{4}}{2}\int dx\psi_{L\downarrow}^{\dagger}(x)\psi_{L\downarrow}(x)\psi_{L\downarrow}^{\dagger}(x)\psi_{L\downarrow}(x)\nonumber \\
 & + & g_{2}\int dx\psi_{R\uparrow}^{\dagger}(x)\psi_{R\uparrow}(x)\psi_{L\downarrow}^{\dagger}(x)\psi_{L\downarrow}(x),\label{eq:interaction}
\end{eqnarray}
where $g_{2(4)\ }$represents the forward scattering strength of different (identical) species. These terms come directly from short range density-density interactions and have been extensively studied in Refs. \cite{Wu2006,Xu2006}. As done previously\cite{Wu2006,Xu2006,Kim2009}, the interacting system can be explored within a Luttinger liquid framework  by representing the fermion fields in terms of  boson fields $\phi$ and $\theta$: $\psi_{R\uparrow}(x)\sim e^{-i(\phi(x)-\theta(x))},\ \psi_{L\downarrow}(x)\sim e^{i(\phi(x)+\theta(x))}.$
Thus, the interacting helical liquid described by $H=H_0+H_\mu+H_I$ is mapped into a free boson gas with a Hamiltonian
\begin{equation}
H=\frac{1}{2\pi}\int dx\left[uK(\nabla\theta)^{2}+\frac{u}{K}(\nabla\phi)^{2}+2\mu\nabla\phi-\delta_S\nabla\theta\right],\label{eq:Bosonization}
\end{equation}
where $u=v((1+\frac{g_{4}}{2\pi v})^{2}-(\frac{g_{2}}{2\pi v})^{2})^{1/2}$
is the renormalized velocity and $K=\left(\frac{1+\frac{g_{4}}{2\pi v}-\frac{g_{2}}{2\pi v}}{1+\frac{g_{4}}{2\pi v}+\frac{g_{2}}{2\pi v}}\right)^{1/2}\ $ is the Luttinger parameter. Values of $K<(>)1\ $represent repulsive (attractive)
interactions. The chemical potential terms $\mu$ and $\delta_S$ can be absorbed as inhomogeneous shifts of the bosonic fields
\begin{eqnarray}
\tilde{\phi}(x)=\phi(x)+\mu Kx/u,\;\;\;\; \tilde{\theta}(x)=\theta(x)-\delta_{S} x/2Ku,
\label{eq:shift}
\end{eqnarray} 
which transforms the Hamiltonian to the standard form $H=\frac{1}{2\pi}\int dx(uK(\nabla\tilde{\theta})^{2}+\frac{u}{K}(\nabla\tilde{\phi})^{2}).$
Thus, while the QSH system bears key differences in the physics, at the technical level, several of its properties can be mapped to the extensively analyzed Luttinger liquid system describing the low-energy physics of a {\it spinless} interacting 1DEG.  

\begin{figure}[t]
\centering{}\includegraphics[width=0.5\textwidth]{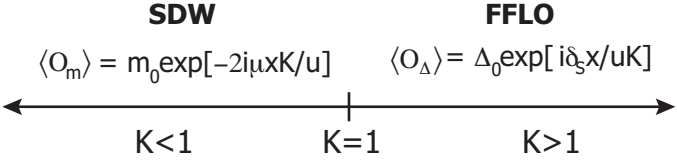}\caption{\label{fig:Phase-diagram}$T=0$ phase diagram of QSH edge  for interactions characterized by $K$, finite chemical potential $\mu$ and finite spin imbalance $\delta_S$.}
\end{figure}

From Eq.~(\ref{eq:shift}), it immediately follows that the magnetic and superconducting orders are both associated with oscillations that are renormalized by the interactions. By noting that $O_{m}  \sim e^{2i\phi(x)},O_{\Delta}  \sim e^{2i\theta(x)}$ and using the shifted forms $O_{m}\sim e^{-2i\mu Kx/u}\tilde{O}_{m}, O_{\Delta}\sim e^{i\delta_Sx/(uK)}\tilde{O}_{m}$, we conclude that
\begin{equation}
\langle O_{m}\rangle=m_{0}\exp\left[i q_m x\right], \ \ \langle O_\Delta\rangle=\Delta_{0}\exp\left[iq_{\Delta}x\right]\label{eq:oscillation}
\end{equation}\noindent where  $q_m =-2\mu K/u$ and $q_{\Delta}=\delta_{S}/uK.$
To determine which of the orders dominates, we inspect the form of the associated susceptibilities, given by $\chi_{m/ \Delta}(x,\tau)=-\langle T_{\tau}O_{m\Delta}(x,\tau)O_{m\Delta}^{\dagger}(0,0)\rangle$, where $\tau\ $is imaginary time. We adapt the standard Luttinger liquid treatment\cite{Giamarchi} to our situation to obtain the following temperature dependence in the Fourier domain:
\begin{eqnarray}
\chi_{m}(k=q_m,\omega=0)&\sim& T^{2K-2}, \nonumber \\ \chi_{\Delta}(k=q_{\Delta},\omega=0)&\sim& T^{(2K^{-1}-2)}.
\label{eq:chi}
\end{eqnarray}

The finite wave-vector dependence reflects the oscillatory behavior in Eq.~(\ref{eq:oscillation}) and the stability of a particular order is indicated by the divergence of the associated susceptibility for $T\rightarrow 0$, as summarized in the phase diagram of Fig. \ref{fig:Phase-diagram}. Hence, for repulsive interactions, $K<1$, the system magnetically orders and is characterized by oscillations whose wave-vector $q_{m}$  is controlled by the applied chemical potential. For attractive interactions, $K>1$, the system tends to form a superconducting state that shows chiral FFLO-type oscillations having the beautiful feature that the wave-vector $q_{\Delta}$ is completely tunable via an applied spin imbalance.
\begin{figure}[t]
 \centering{}\includegraphics[width=0.48\textwidth]{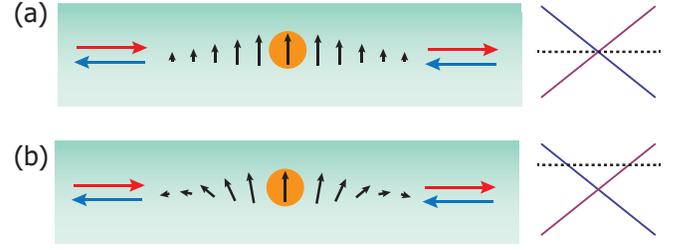}\caption{\label{fig:oscillation}Magnetization oscillations around a single-magnetic impurity on a quantum spin Hall edge which decay as one moves away from the impurity. (a)$\mu=0$ in the weakly repulsive regime leads to a ferromagnetic domain(b) $\mu>0$ in the weakly repulsive regime leads to a domain with oscillatory magnetization direction. }
\end{figure}

Given that the currently available QSH systems are all in the repulsively interacting regime, we now focus on probing the magnetic phase associated with $K<1$.  We show that a weak, localized magnetic impurity that provides an in-plane magnetic field ${\textbf{H}}(x)={\textbf{H}}\delta(x)$ acts as the simplest means of observing the oscillations in the magnetic order of Eq.~(\ref{eq:oscillation}). The coupling to the edge liquid due to such a magnetic perturbation is given by
\begin{eqnarray*}
H_{H} & = & -\mu_{B}\psi^{\dagger}(\sigma_{x}H_{x}+\sigma_{y}H_{y})\delta(x)\psi,\\
 & = & -\mu_{B}|{\bf H}|(O_{m}(x)e^{-i\xi}+O_{m}^{\dagger}(x)e^{i\xi})\delta(x),
\end{eqnarray*}
where $\mu_{B}\ $is the Bohr magneton, $\psi=(\psi_{R\uparrow}\;\;\psi_{L\downarrow})^{T}$
and $\xi=\tan^{-1}(H_y/H_x)$. As shown in Fig. \ref{fig:oscillation}a, a spin up electron impinging the impurity effectively backscatters into a spin down electron and vice-versa. The magnetic perturbation, upon suppressing the spin indices in the $(\psi_{R\uparrow}\;\psi_{L\downarrow})$ fields, exactly maps to the well-known quantum impurity problem in spinless quantum wires whose scaling properties can be easily analyzed within the Luttinger liquid framework\cite{Giamarchi}. In fact, the response to the impurity in our situation parallels the features of  Friedel oscillations in the vicinity of a non-magnetic impurity in a spinless Luttinger\cite{Egger1995}. At high energies and short distances, set by the bare magnetic impurity strength, the impurity can be treated perturbatively. Meanwhile at low energies and large distances, interactions renormalize its strength and the behavior is governed by the strong coupling fixed point wherein the impurity effectively splits the system into two pieces. The resultant magnetization in the helical liquid $m_{+}(x,t)\equiv m_{x}(x,t)+im_{y}(x,t)=2\mu_B\langle O_m(x,t)\rangle$ takes the form
\begin{equation}
m_{+}(x)=\frac{\mu_{B}}{\pi\alpha}e^{i(q_{m}x+\xi)}f(x,T,K,|{\bf H}|),\label{eq:mag}
\end{equation}
where $\alpha$ is a short distance cut-off determined by the bulk energy gap, and $f$ is a dimensionless decaying envelope function whose form depends on the regime being probed~\cite{Egger1996b}. For instance, in the perturbative regime, the  susceptibility of the impurity-free system $\chi_m$ determines the response to the local impurity. For $T\rightarrow 0$, this gives $f\sim x^{1-2K}$ for $\alpha \ll x \ll x_0$, where $x_0$ is a characteristic scale set by the bare impurity strength. On the other hand, for $x\gg x_0$, the strong coupling analysis gives $f\sim x^{-K}$.  For a more general form of magnetic quantum impurity coupling, the helical liquid shows a rich range of behavior, including modified Kondo physics~\cite{Maciejko2009,Tanaka2011} which, in this context, necessitates an investigation of the finite gate-potential induced SDW physics.

Regardless of the strengths of the impurity and interactions, and the regimes being probed, the ubiquitous feature of the magnetization is the $2q_m$ dependence in Eq.~(\ref{eq:mag}) that reflects SDW ordering.  As illustrated in Fig. \ref{fig:oscillation}, the impurity thus creates oscillations in the magnetization which decay with distance.  More explicitly, if for example $H_x\neq 0$ and $H_y=0$ we have $(m_{x},m_{y})\sim(\cos(2\frac{K}{u}\mu x),-\sin(2\frac{K}{u}\mu x)).$ 
To heuristically understand why this oscillation occurs, consider the free system ($K=1$) where for a given a Fermi level $\mu$, the Kramers' pair of states at the Fermi points  are $e^{ik_{F}x}|\uparrow\rangle,\ e^{-ik_{F}x}|\downarrow\rangle$ (Fig. \ref{fig:banddiagram}a).  In this basis, the superpositions $|\pm\rangle = \frac{1}{\sqrt{2L}}(e^{ik_F x}|\uparrow\rangle\pm e^{-ik_F x}|\downarrow\rangle)$ have magnetizations  $m_{+}\sim\pm e^{-2ik_{F}x}$ and  $M_{z}=0.$  A local magnetic field $H_{x}\delta(x)$ breaks the time-reversal symmetry of the pure QSH system, removes the degeneracy between the two states $|+\rangle,\ |-\rangle\ $ since $\langle +\vert H_x\delta(x)\sigma^x\vert +\rangle =-\langle -\vert H_x\delta(x)\sigma^x\vert -\rangle\neq 0,$ and forces an incomplete compensation in magnetization. We emphasize that, in contrast to the oscillations that yield RKKY interactions, the helical nature of the QSH gives rise to spin oscillations in {\it direction} while the magnitude remains fixed. 


In conclusion, we have shown that the helical nature of the QSH is unique in giving rise to ordered  oscillatory phases in the presence of finite chemical potential and spin imbalance. The experimental feasibility of realizing and detecting the SDW phase is promising. For instance, in HgTe quantum wells we expect the interactions to be weakly repulsive and $v\sim 10^5$m/s which leads to a characteristic tunable wavelength of around $200/\mu$ nm where $\mu$ is the chemical potential tuned from the edge state Dirac point in $meV.$ As mentioned, depending on the temperature this oscillation will be modulated by a (perhaps strongly) decaying envelope function.  To detect the oscillations, one could perhaps employ scanning tunneling microscopy as has been successfully
used to observe the RKKY oscillation\cite{Meier2008,RKKY2010} of the magnetization near a magnetic impurity, and   Friedel oscillations\cite{Friedel1954} near a charged impurity\cite{Friedel1993}.  While oscillations in the magnetization direction are harder to detect, any gate-voltage dependent oscillations would be indicative of our proposed SDW phase. The only modification to current setups would be the necessity of a back gate so that the oscillations could be accessed. Finally, the possibility to induce attractive interactions in the QSH system, as has been achieved  in 1D cold atomic gases, would open up the fascinating prospect of realizing the chiral FFLO oscillatory superconducting phase.

\begin{acknowledgments}
We are grateful to W. DeGottardi for his involved discussions. 
This work is supported by the U.S. Department of Energy, Division of Materials Sciences under Award No. DE-FG02-07ER46453 (QLM, TLH, SV), and by the NSF under grant DMR 0758462 (TLH), the AFOSR under grant FA9550-10-1-0459 (MJG) and the ONR under grant N00141110728 (MJG).
\end{acknowledgments}

\end{document}